\documentclass[twoside,twocolumn,9pt]{article}
\usepackage{extsizes}
\usepackage[super,sort&compress,comma]{natbib} 
\usepackage[version=3]{mhchem}
\usepackage[left=1.5cm, right=1.5cm, top=1.785cm, bottom=2.0cm]{geometry}
\usepackage{balance}
\usepackage{mathptmx}
\usepackage{sectsty}
\usepackage{graphicx} 
\usepackage{lastpage}
\usepackage[format=plain,justification=justified,singlelinecheck=false,font={stretch=1.125,small,sf},labelfont=bf,labelsep=space]{caption}
\usepackage{float}
\usepackage{fancyhdr}
\usepackage{fnpos}
\usepackage[english]{babel}
\addto{\captionsenglish}{%
  
}
\usepackage{array}
\usepackage{droidsans}
\usepackage{charter}
\usepackage[T1]{fontenc}
\usepackage[usenames,dvipsnames]{xcolor}
\usepackage{setspace}
\usepackage[compact]{titlesec}
\usepackage{hyperref}

\usepackage{epstopdf}

\definecolor{cream}{RGB}{222,217,201}

\usepackage{tikz}
\usetikzlibrary{external}
\usepackage{amsmath,amsfonts,amssymb,amscd,amsthm,xspace,bm}

\begin{document}

\pagestyle{fancy}
\thispagestyle{plain}
\fancypagestyle{plain}{
\renewcommand{\headrulewidth}{0pt}
}

\makeFNbottom
\makeatletter
\renewcommand\LARGE{\@setfontsize\LARGE{15pt}{17}}
\renewcommand\Large{\@setfontsize\Large{12pt}{14}}
\renewcommand\large{\@setfontsize\large{10pt}{12}}
\renewcommand\footnotesize{\@setfontsize\footnotesize{7pt}{10}}
\makeatother

\renewcommand{\thefootnote}{\fnsymbol{footnote}}
\renewcommand\footnoterule{\vspace*{1pt}%
\color{cream}\hrule width 3.5in height 0.4pt \color{black}\vspace*{5pt}} 
\setcounter{secnumdepth}{5}

\makeatletter 
\renewcommand\@biblabel[1]{#1}            
\renewcommand\@makefntext[1]%
{\noindent\makebox[0pt][r]{\@thefnmark\,}#1}
\makeatother 
\renewcommand{\figurename}{\small{Fig.}~}
\sectionfont{\sffamily\Large}
\subsectionfont{\normalsize}
\subsubsectionfont{\bf}
\setstretch{1.125} 
\setlength{\skip\footins}{0.8cm}
\setlength{\footnotesep}{0.25cm}
\setlength{\jot}{10pt}
\titlespacing*{\section}{0pt}{4pt}{4pt}
\titlespacing*{\subsection}{0pt}{15pt}{1pt}

\fancyfoot{}
\fancyfoot[LO,RE]{\vspace{-7.1pt}\includegraphics[height=9pt]{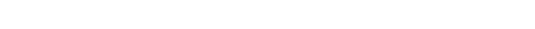}}
\fancyfoot[CO]{\vspace{-7.1pt}\hspace{13.2cm}\includegraphics{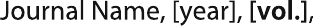}}
\fancyfoot[CE]{\vspace{-7.2pt}\hspace{-14.2cm}\includegraphics{head_foot/RF}}
\fancyfoot[RO]{\footnotesize{\sffamily{1--\pageref{LastPage} ~\textbar  \hspace{2pt}\thepage}}}
\fancyfoot[LE]{\footnotesize{\sffamily{\thepage~\textbar\hspace{3.45cm} 1--\pageref{LastPage}}}}
\fancyhead{}
\renewcommand{\headrulewidth}{0pt} 
\renewcommand{\footrulewidth}{0pt}
\setlength{\arrayrulewidth}{1pt}
\setlength{\columnsep}{6.5mm}
\setlength\bibsep{1pt}

\makeatletter 
\newlength{\figrulesep} 
\setlength{\figrulesep}{0.5\textfloatsep} 

\newcommand{\topfigrule}{\vspace*{-1pt}%
\noindent{\color{cream}\rule[-\figrulesep]{\columnwidth}{1.5pt}} }

\newcommand{\botfigrule}{\vspace*{-2pt}%
\noindent{\color{cream}\rule[\figrulesep]{\columnwidth}{1.5pt}} }

\newcommand{\dblfigrule}{\vspace*{-1pt}%
\noindent{\color{cream}\rule[-\figrulesep]{\textwidth}{1.5pt}} }

\makeatother

\twocolumn[
  \begin{@twocolumnfalse}
{\includegraphics[height=30pt]{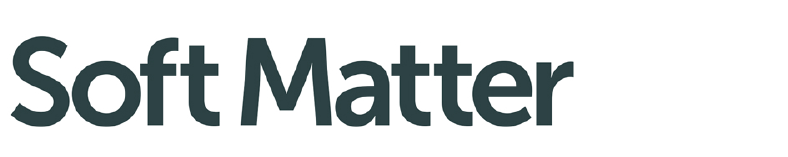}\hfill\raisebox{0pt}[0pt][0pt]{\includegraphics[height=55pt]{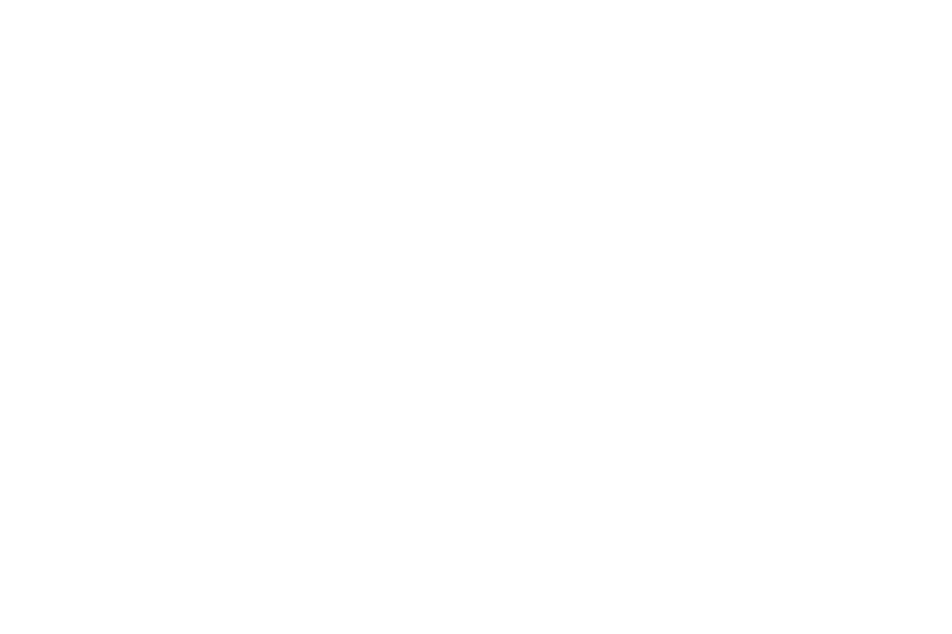}}\\[1ex]
\includegraphics[width=18.5cm]{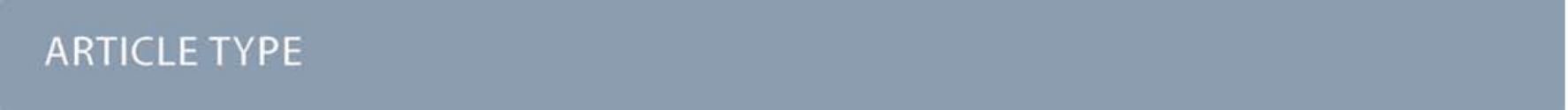}}\par
\vspace{1em}
\sffamily
\begin{tabular}{m{4.5cm} p{13.5cm} }

\includegraphics{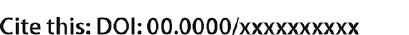} & \noindent\LARGE{\textbf{Capillary interactions between soft capsules protruding through thin fluid films}} \\
\vspace{0.3cm} & \vspace{0.3cm} \\

& \noindent\large{Maarten Wouters\textit{$^{a}$}, Othmane Aouane\textit{$^{b}$}, Marcello Sega\textit{$^{b}$}  and Jens Harting$^{\ast}$\textit{$^{b,c}$}} \\

\includegraphics{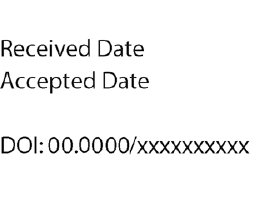} & \noindent\normalsize{When a suspension dries, the suspending fluid evaporates, leaving behind a dry film composed of the suspended particles. During the final stages of drying, the height of the fluid film on the substrate drops below the particle size, inducing local interface deformations that lead to strong capillary interactions among the particles. Although capillary interactions between rigid particles are well studied, much is still to be understood about the behaviour of soft particles and the role of their softness during the final stages of film drying. 
Here, we use our recently-introduced numerical method that couples a fluid described using the lattice Boltzmann approach to a finite element description of deformable objects to investigate the drying process of a film with suspended soft particles. Our measured menisci deformations and lateral capillary forces, which agree well with previous theoretical and experimental works in case of rigid particles, show that the deformations become smaller with increasing particles softness, resulting in weaker lateral interaction forces. At large interparticle distances, the force approaches that of rigid particles. 
Finally, we  investigate the time dependent formation of particle clusters at the late stages of the film drying.   
}\\ 

\end{tabular}

 \end{@twocolumnfalse} \vspace{0.6cm}

  ]

\renewcommand*\rmdefault{bch}\normalfont\upshape
\rmfamily
\section*{}
\vspace{-1cm}


\footnotetext{\textit{$^{a}$~Department of Applied Physics, Eindhoven University of Technology, De Rondom 70, 5612 AP, Eindhoven, The Netherlands; E-mail: m.p.j.wouters@tue.nl}}
\footnotetext{\textit{$^{b}$~Helmholtz Institute Erlangen-N\"{u}rnberg for Renewable Energy, Forschungszentrum J\"{u}lich, F\"{u}rther Stra{\ss}e 248, 90429 N\"{u}rnberg, Germany. }}
\footnotetext{\textit{$^{c}$~Department of Chemical and Biological Engineering and Department of Physics, Friedrich-Alexander-Universit\"at Erlangen-N\"urnberg, F\"{u}rther Stra{\ss}e 248, 90429 N\"{u}rnberg, Germany.}}



\section{Introduction}
Capillary interactions give rise to a wide range of interesting phenomena, and the first documented observations date back to the Renaissance, when Leonardo da Vinci described the capillary rise in a glass tube~\cite{Hardy1922}.
The clustering of Cheerios in a bowl of milk~\cite{Vella2005} and of mosquito eggs on the water surface~\cite{Loudet2011} are all examples of gravity-induced capillary interactions.
Microswimers made of ferromagnetic beads on a fluid interface can be set in motion on the interface surface thanks to the interplay between the capillary forces and a controlled magnetic field~\cite{LaGr16,Sukhov2019}.
The famous coffee-stain effect\cite{Deegan1997} can be suppressed by shape-dependent capillary interactions~\cite{Yunker2011,Mampallil2018}. Furthermore, the assembly of anisotropic particles at a fluid interface can be controlled by use of switchable dipolar capillary interactions~\cite{Davies2014}.

A single spherical particle adsorbed onto a fluid interface of a film thinner than the particle diameter deforms the interface around it symmetrically and experiences in general no forces parallel to the substrate.
The extent over which the particle deforms the meniscus around it is characterised by the capillary length ${\ensuremath{L_{\mathrm{cap}}}}$.
When the separation between two particles is below this characteristic length, the capillary force between the particles becomes relevant, and the driving mechanism for all the phenomena described in the previous paragraph.
Depending on the sign of the slopes of the meniscus at the contact points with the particles, this force can be either attractive or repulsive~\cite{Kralchevsky1994,Kralchevsky2000}, and can be used to direct the self-assembly of particles~\cite{Grzelczak2010,Deshmukh2015,Morris2015,XiDa15}.

\begin{figure}
\begin{center}
    \includegraphics[width=1.\linewidth]{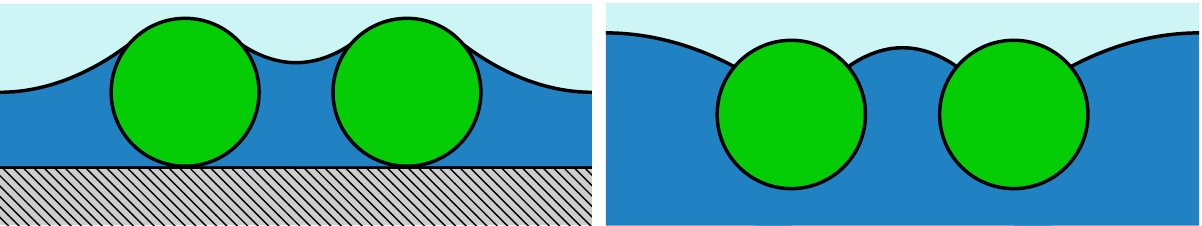}
\end{center}
    \caption[Visualisation of lateral immersion and flotation forces]{Lateral capillary interactions between spherical particles:
        (left) Wetting-dominated interface deformation;
        (right) External force-dominated interface deformation.
    }
    \label{fig:immersion-flotation}
\end{figure}

The lateral capillary interaction forces are referred to as \textit{lateral flotation forces} when the meniscus deforms such that the gravitational potential energy of the particles reduces with the inter-particle separation~\cite{Kralchevsky1994}.
\textit{Lateral immersion forces}, instead, are defined as those appearing when the deformation of the meniscus is related to the wetting properties of the particle surface.
In Fig.~\ref{fig:immersion-flotation} we schematically visualise these two groups.
The strength of lateral flotation forces is proportional to $R^6/\gamma$, while the lateral immersion forces are proportional to $R^2/\gamma$, where $R$ is the particle radius and $\gamma$ is the surface tension~\cite{Kralchevsky1994}.
Due to this difference, capillary interactions between spheres protruding through a thin fluid film can be several orders of magnitude larger than the force between the same spheres floating at a liquid-liquid interface~\cite{Velev1993,Kralchevsky1993,Paunov1993}.

In this paper we focus on lateral immersion forces between particles protruding through a fluid film on top of a rigid, flat substrate.
The capillary interactions between rigid particles in liquid films have already been studied extensively. Theoretical descriptions have been derived for rigid spheres and cylinders for both immersion and flotation forces~\cite{Kralchevsky1993,Kralchevsky1994,Morris2008,Paunov1993}.
Several experimental studies report on the interaction forces between particles, and on the clustering and aggregation of particles due to these capillary interactions~\cite{Danov2010,Dalbe2011,Pieranski1980,Bowden2001}.
Additionally, analytical and computational techniques have been applied to study the capillary interactions between rigid particles~\cite{Connington2015,Onishi2008,Stratford2005_2,Davies2014,Liang2013,Nishikawa2003,Fujita2007,Guzowski2019,XiDa15,Dominguez2010,Oettel2005}.

Several experimental studies for soft latex or microgel particles can be found~\cite{Velikov1998,Kooij2015a,Rey2020,Brugger2008,Deshmukh2015}. 
Furthermore, molecular dynamics, dissipative particle dynamics, lattice Boltzmann and mean field simulations were applied to investigate various properties of soft particles at fluid interfaces~\cite{Mehrabian2016,arismendi-arrieta_deformability_2020,doukas_structure_2018,yong_modeling_2015,bushuev_compression_2020,paper1}. 
While molecular dynamics simulations are not capable to study the influence of the particle softness and the surface wetting properties for many particles at a fluid-fluid interface due to the prohibitive computational effort, dissipative particle dynamics and mean field approaches overcome this limitation by reducing the resolved details of the particle structure. However, to the best of our knowledge,  
the detailed influence of the particle softness on the lateral forces and on aggregation properties has not been systematically studied in experiments or simulations.

The remainder of the paper is organised as follows.
In section~\ref{sec:theory} we provide a brief summary of the relevant theory on lateral immersion forces between rigid spherical particles.
Section~\ref{sec:method} then gives a brief overview of the used numerical method and simulation set up.
In section~\ref{sec:results}, we benchmark our method and compare simulations of rigid spherical particles with both theory and experimental results.
Next, we characterise the deformation of the interface and capillary charge of a single soft particle in a liquid film on a rigid substrate, and study the influence of the softness and wetting properties of the particle.
Furthermore, the clustering of many soft particles in a liquid film is studied.
Finally, we discuss our results and present our conclusions.

\section{Lateral capillary interactions}\label{sec:theory}
Here, following Kralchevsky and coworkers~\cite{Kralchevsky2000,Kralchevsky2001_book}, we summarize the theory of the capillary interaction between two spheres with radius $\ensuremath{R_0}$ protruding through a fluid layer of height $\ensuremath{h_0}$ above a solid, flat substrate as depicted in the left panel of Fig.~\ref{fig:immersion-flotation}.

We define the height of the fluid meniscus in the horizontal \textit{xy}-plane relative to the film height at infinite distance as $\ensuremath{h_0}$.

Sufficiently far away from the particle the meniscus is flat, and its overall shape can be described by the Laplace equation of capillarity\cite{Kralchevsky1993,Princen1969}
\begin{equation}
    \ensuremath{\gamma} \nabla_{{\mathrm{II}}} \cdot \bigg[\frac{\nabla_{{\mathrm{II}}}\ensuremath{\zeta}}{(1+|\nabla_{{\mathrm{II}}}|^2)^{1/2})}\bigg] = P_c,
\end{equation}
where \ensuremath{\gamma} is the surface tension, \ensuremath{\zeta} describes the height of the meniscus relative to $\ensuremath{h_0}$ in the horizontal $xy$-plane, $P_c$ is the capillary pressure difference across the meniscus, and
\begin{equation}
    \nabla_{{\mathrm{II}}} = \vec{e}_x \frac{\partial}{\partial x} + \vec{e}_y \frac{\partial}{\partial y},
\end{equation}
is the two-dimensional gradient operator in the $xy$-plane.
If the particle deforms the interface only slightly, with pertubations that are small as compared to the undisturbed film thickness $\ensuremath{h_0}$, the Laplace equation simplifies to a linear form~\cite{Kralchevsky2000}
\begin{align}
    \nabla_{{\mathrm{II}}}^2\ensuremath{\zeta} = \ensuremath{q}^2 \ensuremath{\zeta},
    \label{eq:grad-q}
\end{align}
where \ensuremath{q} is the inverse capillary length, which characterises the extent of the deformation of the fluid meniscus.
Converting Eq.~\ref{eq:grad-q} to cylindrical coordinates $(r, \theta)$, the meniscus shape around a single particle can be shown to take the form~\cite{Kralchevsky1993,Kralchevsky2001}
\begin{align}
    \ensuremath{\zeta}(r) 
        &=\ensuremath{\Gamma}\sin(\ensuremath{\psi}) K_0(\ensuremath{q}r),
    \label{eq:meniscus}
\end{align}
where \ensuremath{\Gamma} is the radius of the three-phase contact line, \ensuremath{\psi} is the slope of the meniscus, and $K_0$ is the modified Bessel function of the second kind (Macdonald function~\cite{Korn1968}) and zeroth order,
\begin{equation}
    K_0(x) = \int\limits_0^\infty \frac{\cos(x t)}{\sqrt{t^2+1}} dt.
\end{equation}
The corresponding meniscus decays exponentially at infinity~\cite{Princen1969}.

The capillary interaction force between two spheres partially immersed in a thin film can be approximated by the interaction between two cylinders assuming that the surface tension of the fluids acts only at the particle-fluid contact line and that the curvature of the particle is small as compared to the deformation of the fluid interface~\cite{Kralchevsky1993}.

Despite that the capillary force acting on two immersed bodies results indirectly from the overlap of their menisci, the interaction forces do obey Newton's third law~\cite{Kralchevsky1993,Kralchevsky1994}.
Hence, it suffices to characterise the capillary interaction force on only one of the two bodies.
The original theory is derived in two alternative manners: an energetical approach~\cite{Paunov1993} and a mechanical approach~\cite{Kralchevsky1993,Paunov1992}, which were shown to yield equivalent results~\cite{Kralchevsky1994}.

For the case of two vertical cylinders, labelled as \ensuremath{k}=1,2, the deformation of the interface can be characterised by their capillary charge~\cite{Paunov1993}
\begin{equation}
    \ensuremath{Q}_\ensuremath{k} = \ensuremath{\Gamma}_\ensuremath{k} \sin(\ensuremath{\psi}_\ensuremath{k}),
\end{equation}
where $\ensuremath{\Gamma}_\ensuremath{k}$ is the radius of circumference of the horizontal plane at the contact-point of the fluid meniscus with particle \ensuremath{k}, and \ensuremath{\psi} is the slope of the meniscus near the contact point of particle \ensuremath{k}.
The resulting lateral capillary interaction force reads~\cite{Paunov1993,Velev1993}
\begin{equation}
    F_\mathrm{cap}(L) = -2\pi\ensuremath{q}\ensuremath{\gamma} \ensuremath{Q}_1 \ensuremath{Q}_2 K_1(\ensuremath{q}L),
    \label{eq:interaction-force}
\end{equation}
where $L$ is the distance between the centre of mass of the two particles, and $K_1(x)$ is the modified Bessel function of the second kind and first order
\begin{equation}
    K_1(x) = \frac{\sqrt{\pi}}{(-\frac{1}{2})!} \bigg(\frac{1}{2}x\bigg) \int\limits_1^\infty e^{-xt} (t^2-1)^{-1/2}dt.
\end{equation}
When \ensuremath{q}$L\ll1$ this reduces to a form similar to Coulomb's law for the electric force~\cite{Paunov1993}
\begin{equation}
    F_\mathrm{cap}(L) = -2\pi\ensuremath{\gamma}\frac{\ensuremath{Q}_1 \ensuremath{Q}_2}{L},
\end{equation}
which is why $\ensuremath{Q}_\ensuremath{k}$ is commonly referred to as the capillary charge.

Although derived initially only for contact angles close to $\pi/2$, Velev~{\it {et al}}~\cite{Velev1993} have shown that it remains valid even for highly wetting surfaces, and thus for large meniscus slopes in the vicinity of the cylinders.

\section{Numerical method}\label{sec:method}
\subsection{Method description}
We simulate the suspending fluid using the lattice Boltzmann method (LBM)~\cite{Succi2001}. 
The standard LBM can be extended towards multiphase/multicomponent fluids~\cite{Shan1993,Liu2016} and suspensions of particles of arbitrary shape and wettability~\cite{ladd-verberg2001,Jansen2011,XiDa15}. {We review some details in the following and refer the reader to Ref.\citenum{paper1} for a detailed description of the method and our implementation.}

We solve the discretized Boltzmann transport equation on a cubic lattice with lattice constant $\Delta x$ for the distribution functions of each component $c$,
\begin{equation}
   \!\!\!\! f_i^c(\mathbf{x} + \mathbf{e}_i \Delta t , t + \Delta t)-f_i^c(\mathbf{x},t) = \frac{-\Delta t} {\tau^c} \bigg[  f_i^c(\mathbf{x},t) -f_i^\mathrm{eq}(\mathbf{x},t)\bigg]
    \mbox{,}
    \label{eq:method:boltzmann}
\end{equation}
where $i=1,...,19$ labels the discrete velocity vectors in three dimensions, $f_i^c(\mathbf{x},t)$ is the single-particle distribution function, $\Delta t$ is the time step, and $\mathbf{e}_i$ is the discrete velocity in the $i$th direction. Here, $\tau^c$ represents the relaxation time for component $c$. We define the macroscopic densities and velocities for each component as $ \rho^c(\mathbf{x},t) = \rho_0
\sum_if^c_i(\mathbf{x},t)$, where $\rho_0$ is a reference density, and
$\mathbf{u}^c(\mathbf{x},t) = \sum_i  f^c_i(\mathbf{x},t)
\mathbf{e}_i/\rho^c(\mathbf{x},t)$, respectively. 
$f_i^\mathrm{eq}$ is the second-order equilibrium distribution function, defined as
\begin{equation}
  \label{eq:lbm:equilibrium}
  f_i^{\mathrm{eq}} = \omega_i \rho^c \bigg[ 1 + \frac{\mathbf{e}_i \cdot \mathbf{u}^c}{c_s^2} - \frac{ \left( \mathbf{u}^c \cdot \mathbf{u}^c \right) }{2 c_s^2} + 
  \frac{ \left( \mathbf{e}_i \cdot \mathbf{u}^c \right)^2}{2 c_s^4}  \bigg]
  \mbox{,}
\end{equation}
where $\omega_i$ denotes the lattice weights with values $\omega_0=1/3$ for the rest component, $\omega_{1,\dots,6}=1/18$ for the six nearest neighbors and $\omega_{7,\dots,18}=1/36$ for the nearest neighbours in diagonal direction.
The speed of sound of the model is $c_s = \frac{1}{\sqrt{3}} \frac{\Delta x}{\Delta t}$.

The polymeric soft particles are modelled using fluid-filled elastic capsules~\cite{lac2004spherical}. We use a strain-hardening two-dimensional hyperelastic law known as the Skalak strain energy~\cite{skalak1973strain}, which is written as 
\begin{equation}
    E^\textnormal{strain} = \frac{\kappa_S}{4} \oint (I_1^2 +2I_1 - 2I_2 + C I_2^2) dA_c, \quad C > -1/2,
    \label{eq:method:strain}
\end{equation}
where $\oint$ is an integral over the capsule area ($A_c$), $I_1=\lambda_1^2 + \lambda_2^2 -2$ and $I_2 = \lambda_1^2 \lambda_2^2 -1$ are the deformation invariants, and $C$ is a constant parameter related to the strain-hardening nature of the membrane. 
In the small deformation limit, the 2D Poisson ratio can be expressed as function of $C$ such as $\nu_s = C/(1 +C)$ with $\nu_s\in ] - 1\ldots 1]$~\cite{Barthes2016}.
The area dilatation modulus {$\kappa_A$} is defined such as  {$\kappa_A/\kappa_S= 1 + 2C$}.
{To avoid membrane buckling, which can occur as a result of compressive tensions \cite{lac2004spherical}, we restrict ourself to quasi-inextensible membranes with $C \approx 9$.}
In addition to resistances to shear elasticity and area dilatation, our particles are endowed with bending resistance. The curvature energy is accounted for via the Helfrich free energy
\begin{equation}
    {E}^\textnormal{bending} = \frac{\kappa_B}{2}\oint [2H - H_0]^2 dA_c + {\kappa_G}\oint_A K dA_c,
    \label{eq:method:curvature}
\end{equation}
where $H_0$, $H = \frac{1}{2}\sum_{i=1}^2 C_i$, and $K=\prod_{i=1}^2 C_i$ are the spontaneous, mean, and Gaussian curvatures. $\kappa_B$, and $\kappa_G$ are the bending and Gaussian curvature moduli. The volume conservation of the capsule is enforced using a penalty function reading as
\begin{equation}
{E}^\textnormal{volume} = \frac{\kappa_V}{2}\frac{[V-V_0]^2}{V_0},
 \label{eq:vol_cons}
\end{equation}
where $V_0$ is the reference volume of the stress-free capsule, and $\kappa_V$ is a constant parameter. The strain and {volume} forces are evaluated using the principle of virtual work while the curvature force is obtained from the functional derivative of the Helfrich free energy. Further details on the method can be found in \cite{kruger2011efficient,farutin20143d}. 
{In principle, the approach presented here could be extended to model solid elastic particles by considering a tetrahedralized volume mesh coupled with a 3D hyperelastic constitutive law as recently described in \cite{muller2020hyperelastic}.}

{
The equilibrium shape of the particle steams from the interplay between the property of the interface and the elasticity of the membrane. Thus, we introduce the dimensionless number $\beta$ to describe the softness of the particle at the fluid-fluid interface, such that 
\begin{equation}
    \beta = \frac{R_0^2\gamma}{\kappa_B},
\end{equation}
where $\gamma$ is the surface tension of the fluid-fluid interface, and $R_0$ is the radius of the undeformed particle. $\beta$ here is defined with respect to $\kappa_B$, since the contribution of $\kappa_S$ to the equilibrium shape of the particle is found to be negligible in the absence of in-plane forces.
}

{The particle membrane and the fluid are coupled using the half-way bounce-back algorithm with the coupling proposed by Ladd~\cite{Ladd1994_1} as already used for the simulation of soft particle suspensions in single-component fluids~\cite{MacMeccan2009,Reasor2011,Clausen2010}, and a first-order accurate time-integration scheme where each boundary node ($\hat{\mathbf{x}}_i$) is advected in time such that
\begin{equation}
    \hat{\mathbf{x}}_i(t+\Delta t) = \hat{\mathbf{x}}_i(t) + \Delta t\frac{\mathbf{F}_{i}^\textnormal{tot}}{m_i},
\end{equation}
with $m_i$ being the mass of the $i$th boundary node and $\mathbf{F}_{i}^\textnormal{tot}=\mathbf{F}_{i}^\textnormal{bending} + \mathbf{F}_{i}^\textnormal{strain}+\mathbf{F}_{i}^\textnormal{volume} + \mathbf{F}_{i}^\textnormal{int}$ the total membrane force. Here, $\mathbf{F}_{i}^\textnormal{int} = \mathbf{F}_i^{\textnormal{PP}} + \mathbf{F}_i^{\textnormal{PS}}$ refers to a short range Hertz repulsive force added to avoid particle-particle and particle-substrate overlap.
}

{
The coupling of quantities between the boundary elements and boundary nodes is done via a homogeneous scheme where the three nodes of only the corresponding boundary element are given the same weighting factor. 
The half-way bounce back method is known to suffer from so-called staggered momenta which can be prevented by spreading the total exchanged momentum homogeneously over two consecutive time steps~\cite{Ladd1994_1}.
}

{
To include multi-component fluid interactions, we follow the work of Shan and Chen~\cite{Shan1993} and apply a mean-field force $\mathbf{F}^c$ to the fluid components $c$ and $c'$,
\begin{equation}
    \mathbf{F}^\textnormal{SC}(\mathbf{x}, t) = -\psi^c(\mathbf{x}, t) \sum\limits_{c'}G^{cc'}\sum\limits_i \omega_i \psi^{c'}(\mathbf{x}+\mathbf{e}_i, t) \mathbf{e}_i,
    \label{eq:method:shanchen}
\end{equation}
where $\psi^c$ is a pseudo-potential, and $G^{cc'}$ is the fluid interaction strength related to the surface tension.  For the current work we choose $\psi^c = 1- \exp(-\rho^c/\rho_0)$ and limit ourselves to two components. 
}

 {
In addition to the half-way bounce back conditions, we also decouple the interior and exterior fluid interaction forces. To satisfy continuity close to the
boundary, we interpolate the densities for a layer of fluid nodes just outside
our boundary for the inner fluid, and a layer of fluid nodes just inside the
boundary when calculating the interaction force on the fluid nodes just
outside the boundary.  Momentum is conserved since we apply the resulting force that
would act on the fluid node across the boundary to the boundary
element that separates the set of nodes.
This also allows to tune the contact angle of the particle surface by adding an
offset to the interpolated densities as originally proposed by Jansen et al.~\cite{Jansen2011}. However, in this paper we restrict ourselves to neutrally wetting particles.
For more details and a validation of the algorithm please refer to Ref.~\cite{paper1}.
}

\subsection{Simulation setup}
Throughout this work we use two fluids of equal density.
As a result, we cannot a-priori calculate the capillary length of the cylinder, as one could when there is a density difference between the fluids via~\cite{Velev1993}
\begin{equation}
    \ensuremath{q} = \bigg(\frac{|\rho_2 - \rho_1| g}{\ensuremath{\gamma}} \bigg)^{1/2},
    \label{eq:caplength}
\end{equation}
where $\rho_{1,2}$ are the mass densities of the two fluids and $g$ is the gravitational acceleration.
However, a smaller difference between the two fluid component densities increases the capillary length, and thereby reduces the curvature of the fluid-fluid interface.
Hence, our choice of two equal density fluid components is expected to improve the accuracy of the discretization of the interface onto the fluid lattice.
In our simulations we therefore achieve the capillary length not via Eq.~\ref{eq:caplength}, but rather via a fit of the meniscus profile.

Unless specified otherwise, all work is performed on a rigid, spherical particle with radius $\ensuremath{R_0}=10\Delta x$, with 2880 boundary elements. To each boundary node we assign a mass of $m=25\rho_0$.

The particle protrudes through a fluid film of the first component on a solid substrate, as shown in Fig.~\ref{fig:immersion-flotation}a.
The remainder of the system is filled with the second fluid component, and periodic boundary conditions are applied in all directions.
In all simulations we set the fluid-fluid interaction strength $\ensuremath{G^{cc'}}=3.6$, and the initial minority and majority densities of each component to $\ensuremath{\rho_\mathrm{min}^c}=0.04\ensuremath{\rho_0}$ and $\ensuremath{\rho_\mathrm{maj}^c}=0.7\ensuremath{\rho_0}$. 
In order to enhance the equilibration of the diffuse interface, we initialise a single layer of fluid nodes between the two fluid layers, where both components have a density of 0.3.

 {
A short range Hertz force is used to avoid any overlap between boundary nodes from different interfaces and from the substrate such that
\begin{align}
    & \mathbf{F}^{\text{PP}}    = \Xi \delta^{1.5} \hspace{.5cm} \text{if} \hspace{.25cm} \delta < \delta_0,\\
    & \mathbf{F}^{\text{PS}}    = \Xi \delta^{1.5} \hspace{.5cm} \text{if} \hspace{.25cm} \delta < \delta_0.
    \label{eq:interactions}
\end{align}
The interaction strength is fixed to $\Xi=1.5$ in all presented results, and the cut-off distances to $\delta_{0}=1.5\Delta x$. This choice ensures that there is always at least one fluid node between the particles, and the substrate and the particles.}

\begin{figure}
    \centerline{\includegraphics[width = 1.\linewidth]{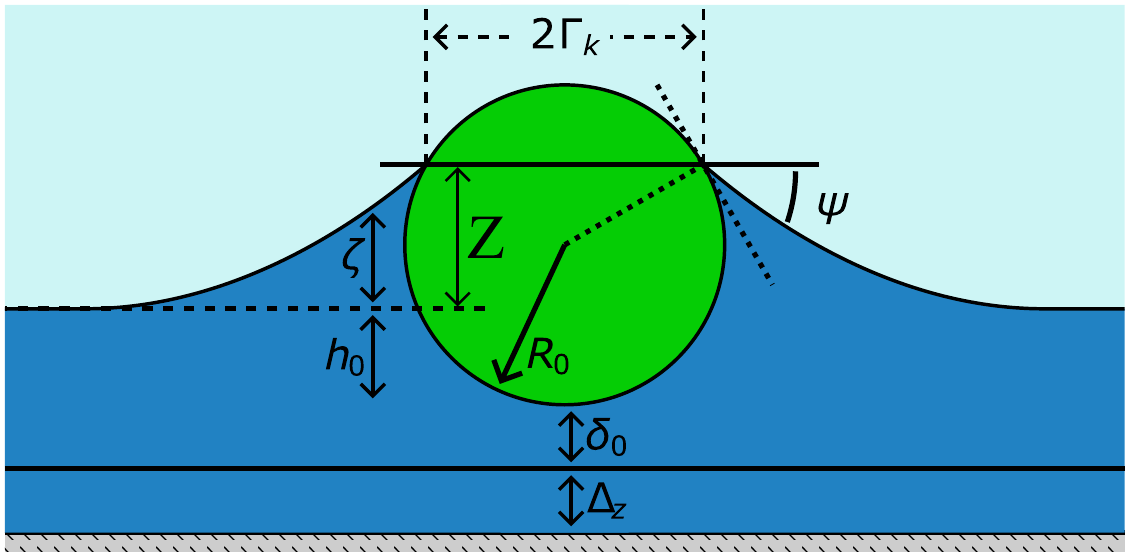}}
    \caption[Visualisation main variables for single particle protruding through a fluid film]{Schematic visualisation of used variables for a single particle (green) immersed in a film of fluid $c$ covered by another fluid $c'$ on top of a substrate (grey).
        The different lengths and sizes are not shown to scale.}
    \label{fig:capillary-interactions:zoom}
\end{figure}

The Shan-Chen multi-component method yields a diffuse interface between the different fluid components.
When a small film is initialised on top of a substrate with a height $\ensuremath{h_0}$ comparable to the diffuse interface width \ensuremath{\xi}, it can be expected that either the flat interface will dewet the solid substrate, or that an additional and undesired interaction appears between the substrate and the interface.
With the fluid-structure coupling that we use, it was shown that for a particle radius of $\ensuremath{R_0} \gtrsim 1.5\ensuremath{\xi}$ the presence of the diffuse interface does not play a significant role on the deformation of a particle~\cite{paper1}.
Still, in order to mitigate possible spurious effects originating from the interaction of the diffuse interface and the substrate, we keep the particles further away from the substrate by introducing an additional  horizontal, repulsive plane with vertical offset $\ensuremath{\Delta_z}=10\ensuremath{\Delta x}$.
A schematic representation of the system, and the main variables is shown in figure~\ref{fig:capillary-interactions:zoom}.

From the perspective of the particle, this repulsive plane acts as the effective location of the substrate, and refer to it simply as to the substrate.

\section{Results}\label{sec:results}

\subsection{Model validation with hard particles}

\begin{figure}
    \centerline{\includegraphics[width = .45\textwidth]{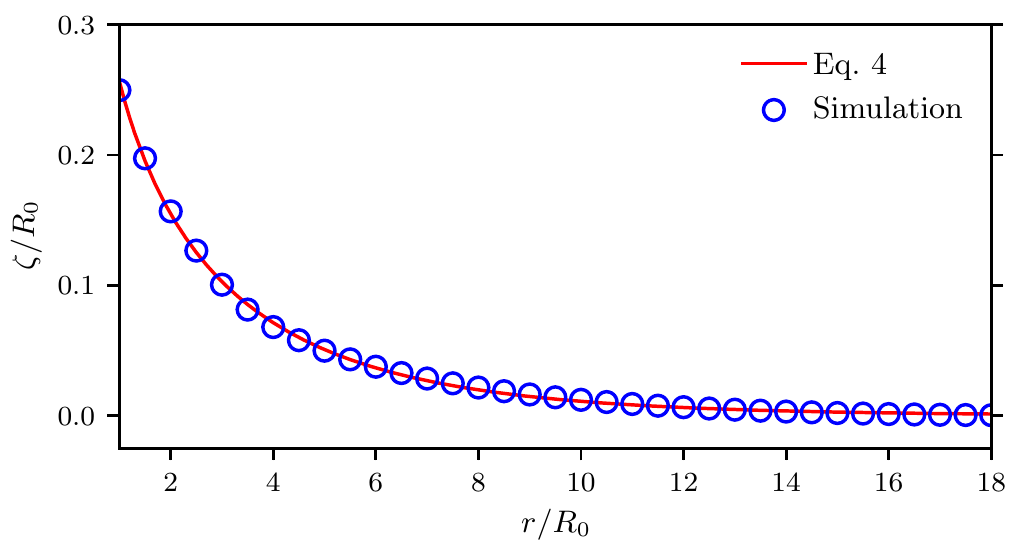}}
    \caption[Equilibrium profile of the fluid meniscus around a rigid particle]{
        Equilibrium profile of the meniscus around a rigid particle immersed in a thin fluid film with $h_0=0.4 R_0$ (circles), and the fitted profile with Eq.~\ref{eq:meniscus} (solid line) for a domain of $\vec{D}=[38.4, 38.4, 3.2]R_0$.
    }
    \label{fig:charge-meniscusfit}
\end{figure}

In order to validate our approach, we begin by performing simulations of a rigid particle. Here, we choose to initialise the fluid film as a flat film, and let the meniscus evolve towards its equilibrium shape. Fig.~\ref{fig:charge-meniscusfit} shows an instantaneous snapshot of the simulated meniscus when the equilibrium has been clearly established ($7.5\times10^5\ensuremath{\Delta t}$), and Eq.~\ref{eq:meniscus} fitted for the inverse capillary length $q$.
The profile is measured on a horizontal slice through the centre of the particle, and averaged over the four sides along the principal directions of the particle. The measured meniscus shape shows an excellent agreement with the theoretical shape, and allows us to obtain a precise estimate of the capillary length and of the capillary charge of the particle.

\begin{figure}
    \centerline{\includegraphics[width = .485\textwidth]{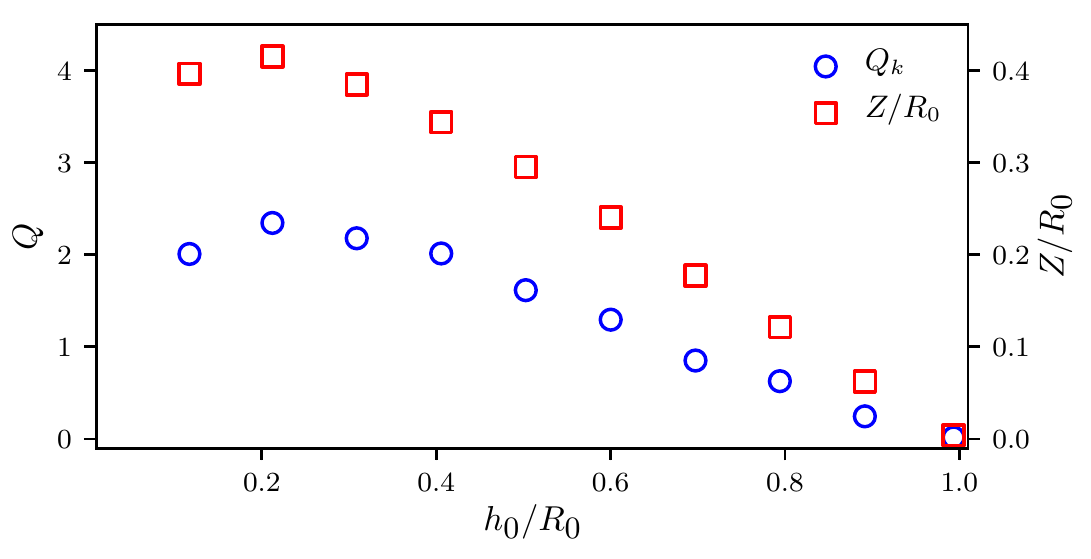}}
    \caption[Capillary charge as function of film thickness]{
        Capillary charge $Q$ (circles) and total rise of the meniscus \ensuremath{Z} (squares) for a single particle immersed in a fluid film for various initial effective film heights $h_0$.
    }
    \label{fig:charge-filmheight}
\end{figure}

Next, we characterise the dependence of the capillary charge of the rigid particle  on the effective height of the film.
Fig.~\ref{fig:charge-filmheight} depicts the measured capillary charge for different particle positions relative to the film height.
When the centre of the particle is close to the fluid interface ($h_0/R_0\simeq 1$), the maximum capillary rise $Z$ and capillary charge $Q$ approach zero, as one can intuitively expect.
For a wide range of relative film heights, both, the rise of the meniscus and the capillary charge appears to follow a linear relation, which only breaks when the fluid interface is getting close to the bottom of the particle, yielding also large capillary charges.

The capillary charge of a single particle determines the interactions with other particles immersed in a fluid film.
After having characterised the dependence of the capillary charge on the main system parameters, we validate the lateral capillary interaction force between two rigid spherical particles immersed in a fluid film with $h_0=0.16 R_0$.
In order to ensure that the fluid meniscus is fully settled when we measure the interaction force, we fix the position of two particles at different separation distances from $L_{gap}=3$ to $7 R_0$.
All simulations are run for $5\times10^5\Delta t$ in order to assure the equilibration of the fluid meniscus, after which data is collected every $10^4\Delta t$.
Since we fix the position of both particles, we can easily extract the total lateral force acting on a particle by sampling the force acting on the boundary nodes that results from fluid-structure coupling.

\begin{figure}
    \centerline{\includegraphics[width = .45\textwidth]{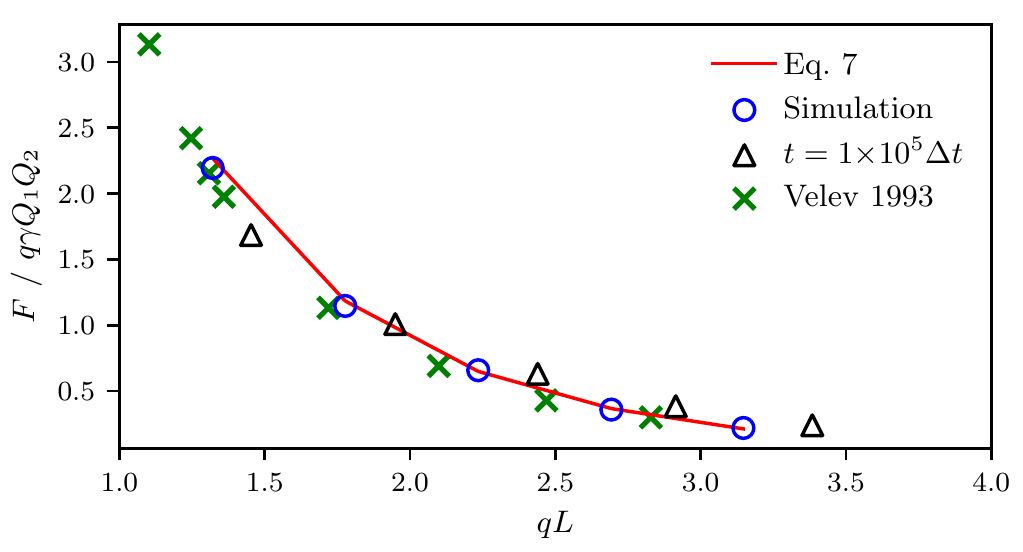}}
    \caption[Validation of lateral capillary interaction force]{
        Effective force between two particles with effective initial film height of $h_0=0.16 R_0$: (cirlces) equilibrium results at $t=5\times10^5 \Delta t$, (upper triangles) results at $t=1\times10^5 \Delta t$, (crosses) experimental results of Velev {\it{et al.}}~\cite{Velev1993},  (solid line) Eq.~\ref{eq:interaction-force} using the capillary charge estimate from the simulation data.
}
    \label{fig:interaction-force}
\end{figure}

In Fig.~\ref{fig:interaction-force} we show the measured lateral interaction force acting between the two particles.
The measured force shows excellent agreement with the theoretical prediction from Eq.~\ref{eq:interaction-force} by taking the capillary charge estimate from the simulation results.
Our simulations show also an excellent agreement with  the experiments of Velev {\it{et al.}} who directly measured the capillary force between two hydrophilic glass cylinders ($\sim300\mu m$ diameter) immersed in a thin film of water~\cite{Velev1993,Kralchevsky2000}.
In this sense, our simulations provide physically sound results, and can be used for studying capillary interactions between particles at fluid-fluid interfaces, provided that the fluid meniscus is allowed to equilibrate for sufficiently long times.

\subsection{Meniscus deformation for soft particles}
We place a deformable particle, initially spherical with radius  $R_0 = 10 \Delta x$, above a thin fluid film, with the lower part of the particle positioned such that it is at the boundary where the interaction with the substrate starts.
The particle softness parameter \ensuremath{\beta} is varied while keeping the Poisson ratio $\nu_s=0.9$ constant.
In contrast to the simulations described previously, the particle is no longer being kept fixed in space, but is left free to move and also to deform.
The fluid-fluid interface is initialised as a flat film, with a final film height of $h_0=0.15 R_0$.

Similar to the rigid particles, the fluid meniscus rises at the particle surface, but at the same time, because of the particle softness, it stretches at the interface. This interplay results in a lowering of the contact-point with the meniscus, and reduces the deformation of the meniscus around the particle.

\begin{figure}
    \centerline{\includegraphics[width = .45\textwidth]{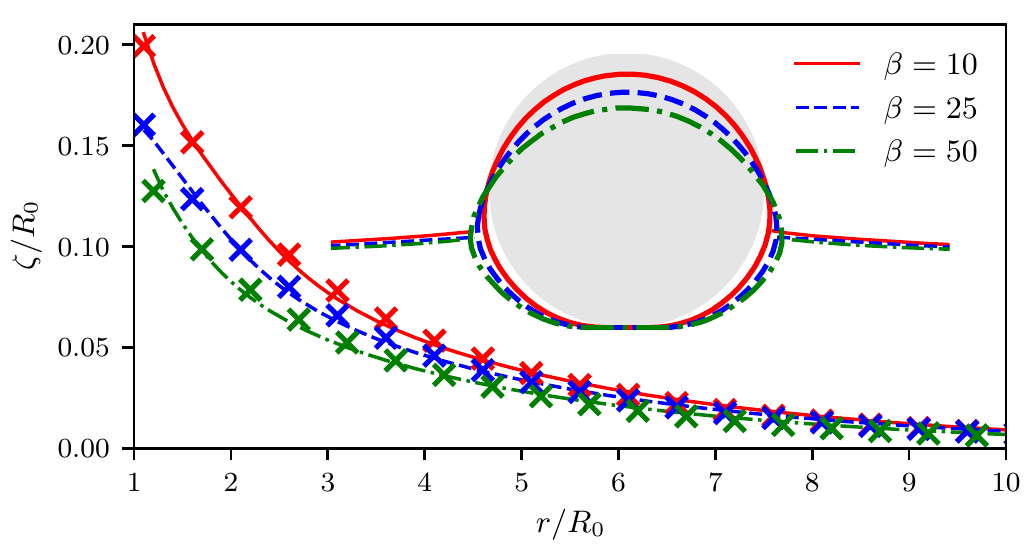}}
    \caption[Fluid meniscus around soft particle for different softness parameters]{
         Radial dependence of the meniscus height for different degrees of softness: (lines) simulation results, (crosses) fit to  Eq.~\ref{eq:meniscus}.
        Inset: density isolines in a slice cut through the centre of the particle (solid lines) and the initial shape of the deformable particle (gray area).
    }
    \label{fig:soft-meniscus}
\end{figure}

In Fig.~\ref{fig:soft-meniscus} we show the averaged height of the fluid meniscus around the particle, compared to the theoretically expected shape as given by Eq.~\ref{eq:meniscus}.
The inset shows the isodensity curves highlighting the equilibrium particle shape and the surrounding meniscus on a slice of the system cut through the center of the particle, for different softness parameters.
The shape of the fluid meniscus resulting from the simulations is in good agreement with the theoretical shape for $\ensuremath{\beta}=10$ and $25$.
However, for $\ensuremath{\beta}=50$, the slope of the meniscus near the contact point is slightly increased as compared to Eq.~\ref{eq:meniscus}.
This offset can be expected to be an error originating from the discretisation of the fluid lattice and the particle boundary near the contact point.

\subsection{Capillary interaction force between two soft particles}
While the slope of the meniscus can be expected to reduce due to the deformation of the soft particle, it is more difficult to predict the changes in contact radius $\Gamma$ and film height as a function of the softness, because of the competition between the reduced rise of the meniscus and the stretching of the particle boundary near the contact points. In general, however, we can expect that the total capillary charge reduces, since the elastic energy stored in the deformation of the soft particle reduces the one stored in the deformation of the fluid-fluid interface.

Computing the non-equilibrium force between two soft particles, free to deform and move is, however, far from being a trivial task, because of possible overlaps of characteristic times for the particles' movement or deformation, and the relaxation time of the meniscus. In order to reduce the inaccuracy in the determination of the force, we opted to let the system start relaxing from a prescribed minimal surface-to-surface distance $L_{\mathrm{gap}}$ of two particles. After $10^4$ timesteps, during which the particle shape does not change significantly anymore, but the interparticle distance is still well below the lattice spacing, we prevent further deformations and let the fluid meniscus relax fully. At the end of this relaxation protocol, we start sampling the force between the two particles.
We then repeat this routine for different initial separations $L_{\mathrm{gap}}$ and softness parameters to characterise the lateral capillary interaction force.
\begin{figure}
    \centerline{\includegraphics[width=0.45\textwidth]{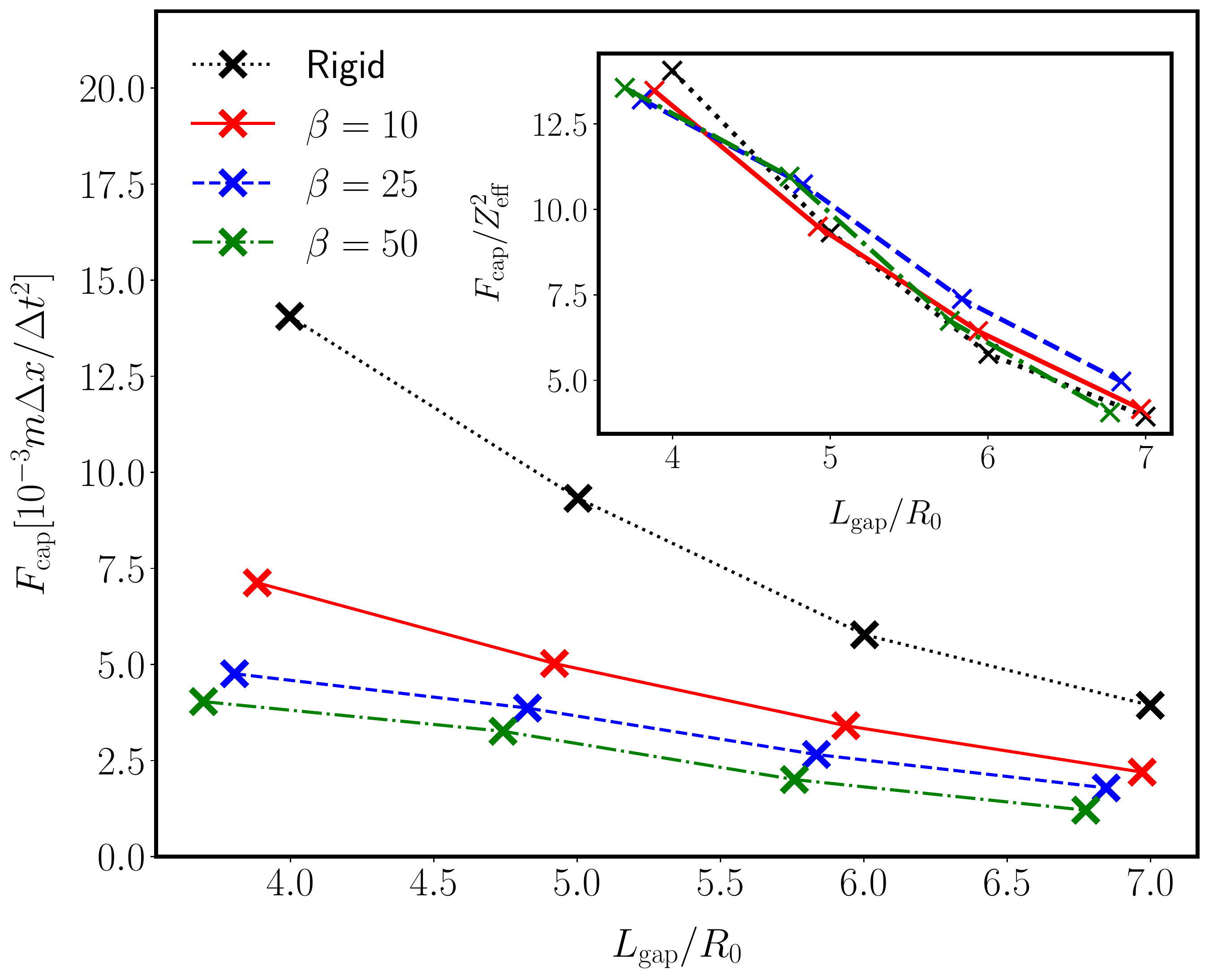}}
    \vspace{-3mm}
    \caption[Capillary interaction force between soft particles]{
         {Capillary force between two particles as a function of the minimal gap between the two: (dotted) rigid, (solid) $\beta=10$, (dashed) $\beta=25$, and (dot-dashed) $\beta=50$.
            The lines are a guide to the eye. Inset: Force scaled by the squared effective maximal rise. }
    }
    \label{fig:interaction-force-soft}
\end{figure}

In Fig.~\ref{fig:interaction-force-soft} we report the measured capillary force/separation curve for different degrees of particle softness.
For all simulated sets of softness parameters we initialise the particles with a gap in between of $L_\mathrm{gap}=4$ to $ 7 R_0$.
Due to the deformation of the particles during the first $10^4$ simulation steps, the gap decreases slightly for the soft particles, while it remains equal for the rigid particles.
As already observed in Fig.~\ref{fig:soft-meniscus}, the fluid meniscus stretches the particle, thereby mitigating the rise of the meniscus at the particle surface.
This results in a reduction of the capillary interaction force for increasing softness parameters as compared to the interaction force of rigid particles at the same particle separation.   {For rigid particles, one can obtain a full collapse of the force/distance curves by rescaling the force with the square of the capillary charge, according to Eq.~\ref{eq:meniscus}. Indeed, we found that such universal behaviour is observable also in the case of deformable particles. In the inset of Fig.~\ref{fig:interaction-force-soft} we show the collapse of the rescaled curves, obtained using capillary charges within the uncertainty margin of the capillary rise.}

\subsection{Approach dynamics of two soft particles}
As shown before, the maximal rise $Z$ of the fluid meniscus at the contact-point with the particle decreases with the particle softness. Therefore, 
the strength of the lateral capillary interaction decreases with the particle softness, too. Qualitatively, one would expect two deformable particles closing the gap in between them more slowly with increasing softness.
\begin{figure}
    \centerline{\includegraphics[width = .45\textwidth]{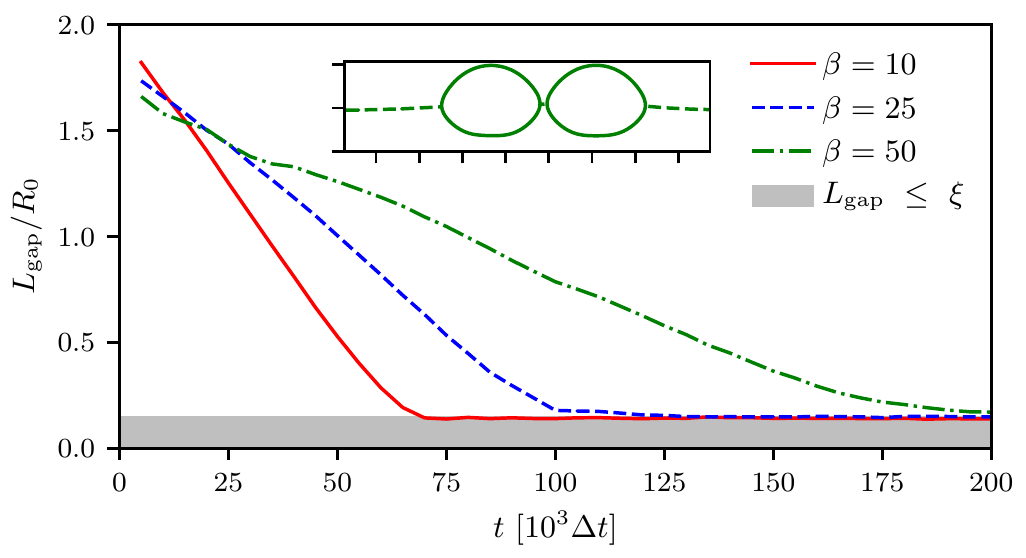}}
    \caption[Capillary interaction force between soft particles]{
        Time evolution of the gap $L_\mathrm{gap}$ between two soft particles, with fluid film height $h_0=0.4 R_0$: (solid line) $\beta=10$, (dashed line) $\beta=25$, (dot-dashed line) {$\beta=50$}.
        The grey area indicates the repulsive region,  $L_\mathrm{gap} < \delta_0$.
        Inset: final state of two particles with $\beta=50$ (solid lines) at the interface (dashed line) 
        Each tick mark in the inset indicates a distance of $R_0$.
            }
    \label{fig:attraction-soft}
\end{figure}

Fig.~\ref{fig:attraction-soft} shows the time evolution of the size of the gap between two soft particles that  start at $L_\mathrm{gap}=2 R_0$ as spherically-shaped particles in a film with $h_0 = 0.4 R_0$.
Indeed, the stiffer particles approach each other faster than the softer particles.
The gap decreases initially as a result of the deformation of the particles and afterwards as a result of the lateral capillary interaction force.
The approaching velocity of the particles is approximately constant until the gap distance is of the order of $\Delta x$. At such close distances, the particle motion is dampened as a result of the hydrodynamic lubrication force between the particles
and finally the repulsive particle-particle interaction force.
For the softest particles (${\ensuremath{\beta}}=50$), the approaching velocity shows some variations due to the changing discretisation of the particle boundary and resulting fluctuations in the force on both the particle boundary and fluid nodes.

\subsection{Clustering of soft particles in a thin film}
The capillary interactions between multiple particles induces, for similarly wetting particle
surfaces, an effective clustering. In this section we study this clustering behaviour in a large
system with O($10^3$) particles, and its dependency on the particle softness.

We initialise $1730$ particles protruding through a fluid film on top of a substrate
corresponding to an effective packing fraction of around 25\% in a domain of $\vec{D}=[1200,1200,40]\ensuremath{\Delta x}^3$.
The initial height of the fluid film is set to $h_0=0.35 R_0$ and the simulations are run for $10^6$ timesteps, while the particle properties are sampled every $10^3$ steps.

\begin{figure*}
    \centering
    \centerline{\includegraphics[width=0.625\textwidth]{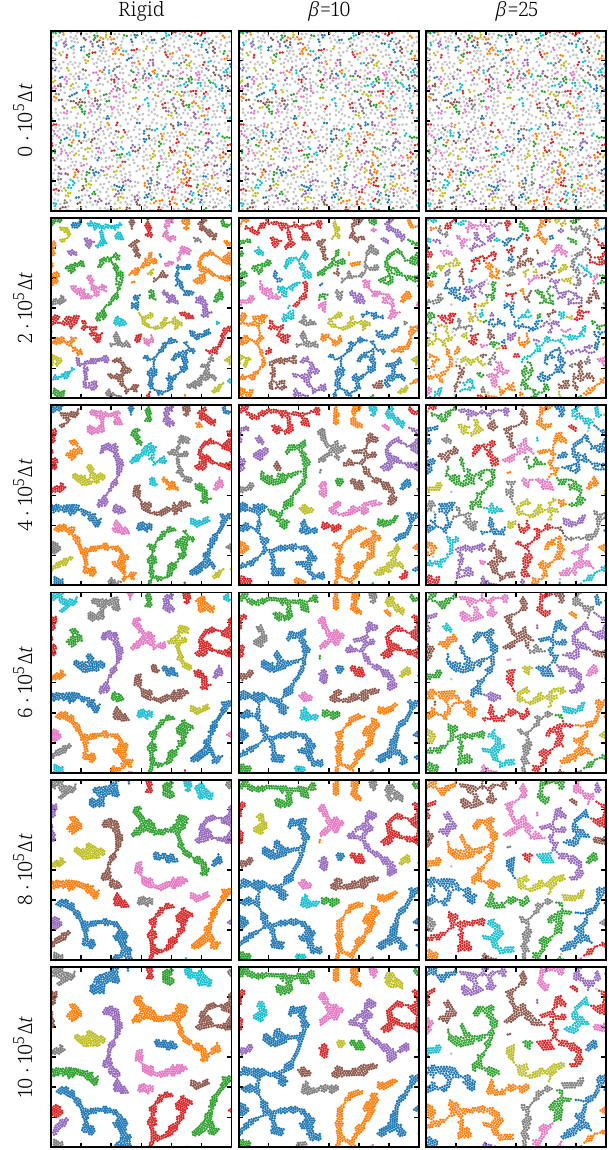}}
    \caption[Clustering in thin film for different softness parameters]{
        Top view of cluster formation in a thin liquid film with 1730 particles protruding the layer.
        The centre of mass of each particle is visualised with a circle, where the colour differentiates between the clusters, and the size is chosen for visualisation purposes.
        Each tick mark indicates a distance of 200$\Delta x$.
    }
    \label{fig:clusters-combined}
\end{figure*}
We initialise the particles at the fluid-fluid interface in a spherical morphology and initialise the fluid-fluid as a flat film on top of the substrate. However,
if the soft particles are not initialised close to their equilibrium shape corresponding to the surrounding fluid interface, the relaxation towards the local equilibrium shape and resulting movement of the particle boundary
induces a flow as well as a deformation and movement of the fluid-fluid interface. The forces acting on the particle due to this relaxation can be substantially stronger than the capillary interactions we are interested in.
This effect is however not easily avoidable, since the equilibrium shapes of the particle and the fluid-fluid interface are not known \textit{a priori}. Therefore, the first few thousand timesteps of the simulations are dominated by the equilibration process until the action of the capillary interactions between the particles becomes the determining factor.

In Fig.~\ref{fig:clusters-combined}, we show some instantaneous snapshots of the particle centres of mass at different times.
The particles are coloured based on the results of a clustering algorithm~\cite{freud}, where particles are grouped into the same group when the separation $\delta_\mathrm{com}$ between their centre of mass satisfies
\begin{equation}
    \delta_\mathrm{com} \leq \langle 2R\rangle + \delta_0.
\end{equation}
Here, $\langle 2R\rangle$ is the particle diameter in the horizontal plane averaged over all particles in the system, and $\delta_0$ is the minimal interaction range between two particle boundary elements.
Due to identical initial conditions, it is possible to appreciate that the morphology of the clusters in the case of rigid particles and in case of $\beta=10$ are quite similar at any stage, whereas the softest case $\beta=25$ is clearly different. 

\begin{figure}
    \centering
    {\includegraphics[width=1.\linewidth]{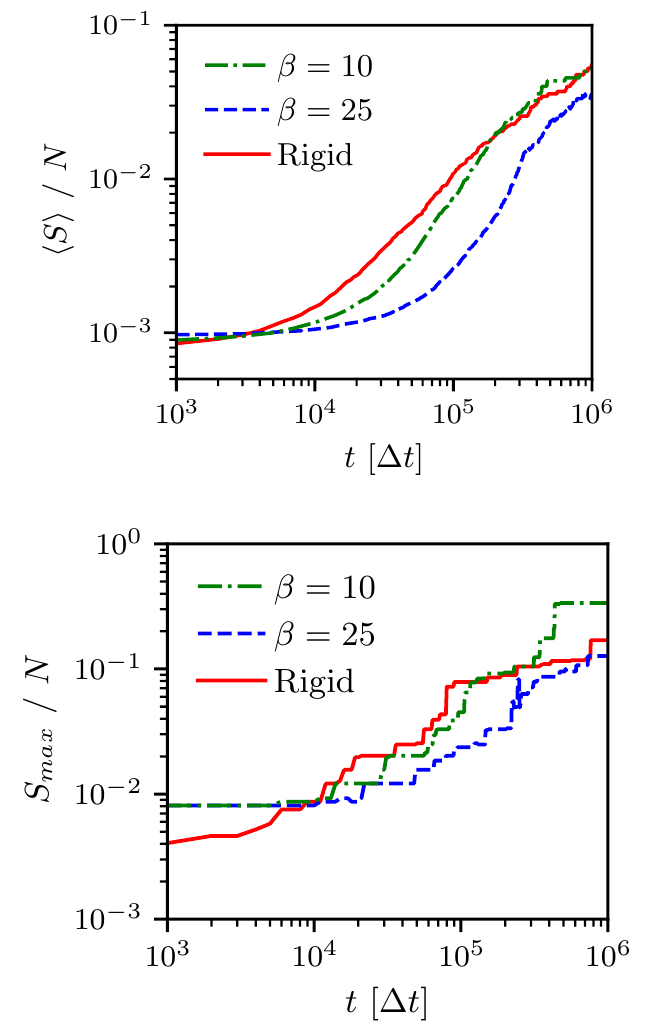}}
    \caption[Evolution of average and maximum cluster size]{
        Evolution of (top) average and (bottom) maximum cluster size (number of particles within the cluster) for different softness parameters: (solid line) rigid, (dot-dashed line) $\beta=10$, (dashed line) $\beta=25$. 
    }
    \label{fig:cluster-properties}
\end{figure}

Looking at the time evolution of the average cluster size, Fig.\ref{fig:cluster-properties}, it appears that the clusters containing rigid particles are growing faster in the initial phase, approximately until $5\times10^4\Delta{}t$, after which the clusters in the system with softness $\beta=10$ start growing faster, being eventually overcome by the softest system, that shows the largest slope at about $5\times10^5\Delta{}t$. 

Initially, we observe a single cluster of six particles, which is a result of the random initialisation of the particles.
Then, in the first stage of the particle clustering, many smaller clusters are rapidly formed as a result of the capillary interactions between the particles.
Here, the rigid particles cluster faster as a result of their larger capillary charge, and the growth rate decreases for an increasing particle softness.

In the second growth stage, roughly between $2.5\times10^3$ and $3\times10^5$ timesteps, 
most particles belong to a cluster, and the average cluster size grows with a constant exponent as can be observed in Fig.~\ref{fig:cluster-properties}. Here, the average domain size
grows faster for the soft particles as compared to the perfectly rigid particles.

The size of the largest cluster is initially the largest for the rigid particles, as a result of the initial fast clustering rate.
However, after roughly $2\times10^5$ steps, the largest cluster for the simulation with particles with \ensuremath{\beta}=10 becomes of a comparable size, and afterwards increases relative to the rigid particles.
The softest particles with $\beta=25$ show a similar trend, albeit somewhat delayed due to the lower initial rate of clustering.

Comparing the different columns of Fig.~\ref{fig:clusters-combined}, one can see that for $\beta=25$ the typical cluster is smaller, but also that the typical separation between the clusters is smaller.
Hence, it can be expected that these clusters are sufficiently close to eventually combine into larger clusters.
Apparently, the initial rapid clustering of the rigid particles results in larger separations between the different clusters, which in the later stage require more effort to cluster together, reducing the rate at which large clusters combine.

\begin{figure}
    \centerline{\includegraphics[width=0.48\textwidth]{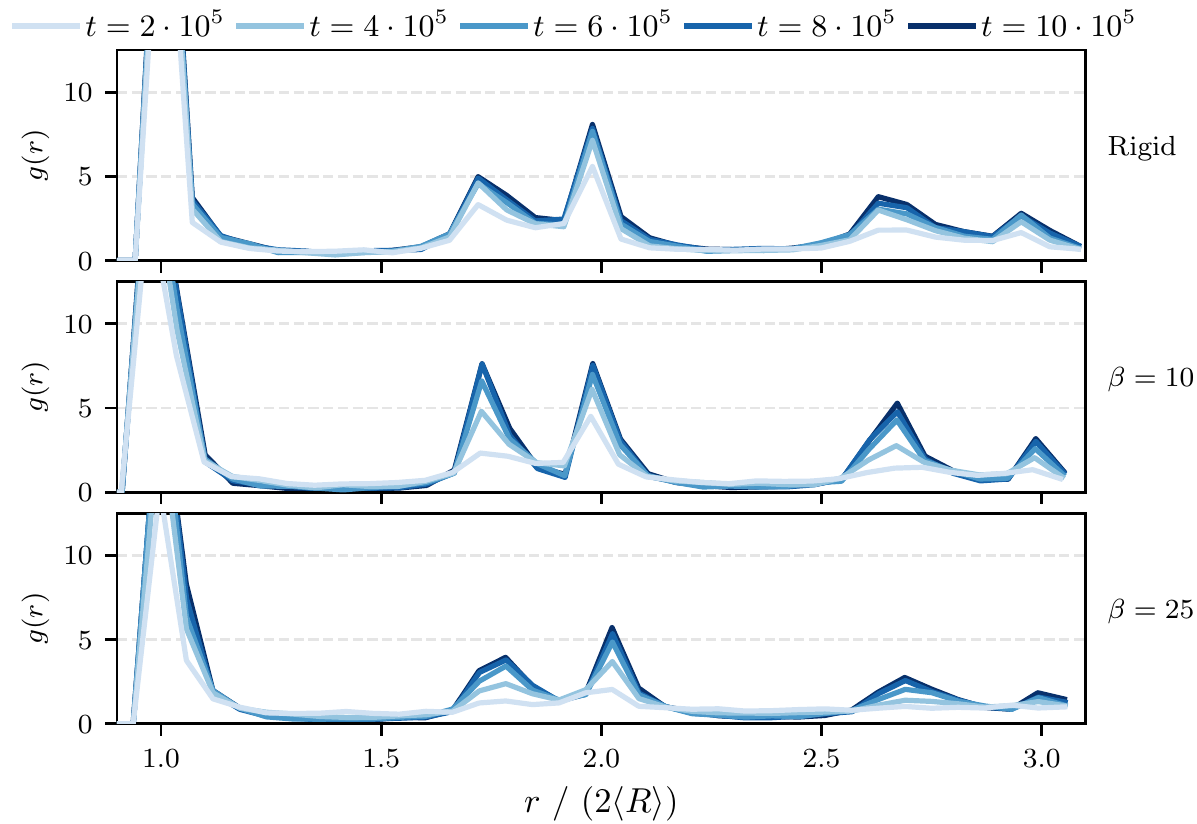}}
    \caption[Radial distribution function for clustering of particles with different softness]{Radial distribution function at different  simulation times. The radial positions are scaled by the instantaneous averaged particle radii in the plane of the fluid-fluid interface.
        The first peak at each plot exceeds the visualised vertical axis, but is not fully shown for clarity purposes.
    }
    \label{fig:cluster-properties-RDF}
\end{figure}

In the early stages the particles are rapidly pulled together by the capillary forces, forming clusters without a uniform order.
For example, in some cases a ring with 7 particles surrounding a single particle or particles oriented in a square packing are formed during the clustering, and remain stable until the end of the simulation.
A thorough visual inspection of Fig.~\ref{fig:clusters-combined} shows the tendency for softer particles to more easily form a hexagonal packing than the rigid particles.
For the rigid particles, however, it occurs more regularly that particles are trapped in a non-hexagonal packing.

Fig.~\ref{fig:cluster-properties-RDF} shows the radial distribution function of the particle centres
\begin{equation}
    g(r)=\frac{A}{N}\frac{n(r)}{2\pi rdr},
\end{equation}
where $n(r)$ is the number of particles within a shell of thickness $dr$ at a radial distance $r$, $A$
is the total area of the system, and $N$ is the total number of particles in the system.
In line with Fig.~\ref{fig:clusters-combined} and~\ref{fig:cluster-properties}, we observe that for the rigid case many particles almost touch each other  already in the early stages (i.e. peaks at integer values occur in $g(r)$), whereas these develop more slowly for softer particles. 
A hexagonal packing (indicated by a peak near $r=\sqrt{3}\langle 2R\rangle$) is more pronounced for $\beta=10$ than for the rigid particles near the end of the simulations.
Furthermore, for the rigid particles most of the regions with a hexagonal packing are formed in the early stages, whereas for softer particles $g(r=\sqrt{3}\langle 2R\rangle)$ increases significantly over time.
In order for clustered particles to transform from an arbitrary packing to a hexagonal packing, they have to move relative to each other.
Soft particles deform near the contact point between particles, which reduces the required energy barrier for particles to move to a hexagonal packing.
Rigid particles are not able to deform, and therefore it is more likely that they remain trapped in a non-hexagonal packing.

\section{Conclusion and discussion}\label{sec:discussion}
We presented three-dimensional numerical simulations of soft  {fluid-filled} particles protruding through a thin fluid film on top of a substrate.
Analytical and experimental results for solid spherical particles were used to validate our model by measuring the meniscus shape and the lateral capillary force. 
In the case of a single soft  {fluid-filled} particle, the deformation of the meniscus decreases with the softness of the particle due to a deformation of the particle induced by the surface tension of the fluid.  {In other words, the capillary charge induced by the particle deforming the interface reduces.}
For two particles, we evaluated the lateral capillary force depending on the particle separation distance. 
At sufficiently large separation distance, the lateral capillary force shows no dependency on the softness of the particles and even converges towards the case of rigid particles.
However, at small separations, the force exhibits a non-monotonic dependency with respect to the distance. We show that our method can also be applied to study capillary force induced clustering of many particles and how the particle softness influences the time dependent size distribution of particle aggregates.

 {As an outlook, future work could include a variation of the particle contact angle. However, while this will certainly have a quantitative effect on our results, we expect the general findings to not change. A non-neutrally wetting soft particle will position and deform asymmetrically with respect to the fluid interface. Thus, the effective capillary charge might be reduced for strongly wetting particles in a thin fluid film.
}

 {
A further possible extension comprises solid elastic particles. For this, the particle model needs to be extended by computing the elastic forces on a tetrahedralized volume mesh rather than on a triangular surface mesh~\cite{muller2020hyperelastic}. However, we do not expect the impact of the actual kind of deformation to change our general conclusions. The particle deforms due to the interplay of the interfacial tension forces acting at the 3-phase contact line, as well as the deformation at its bottom once it gets in contact with the substrate. The deformation at the bottom should be qualitatively similar for elastic particles or soft shells if their effective softness is comparable. At the three-phase contact line, however, a soft shell can show a more smooth curvature as compared to a tip-like deformation observed for soft elastic particles~\cite{Mehrabian2016}. We speculate that this might have an effect on the magnitude of the capillary interactions due to a different resulting interface deformation.
}

\section*{Conflicts of interest}
There are no conflicts to declare.

\section*{Acknowledgements}
We acknowledge financial support from the Dutch Research Council NWO/TTW (project 10018605), HPC-Europa3 (Grant INFRAIA-2016-1-730897), and the German Research Foundation DFG (research unit FOR2688, project HA4382/8-1).
The authors thank the J\"ulich Supercomputing Centre and the High Performance Computing Centre Stuttgart for the allocated CPU time.



\balance



\providecommand*{\mcitethebibliography}{\thebibliography}
\csname @ifundefined\endcsname{endmcitethebibliography}
{\let\endmcitethebibliography\endthebibliography}{}

\end{document}